\documentclass{article}

\usepackage{hyperref}
\usepackage{cite}
\usepackage{amsmath,amssymb,amsfonts}
\usepackage{algorithmic}
\usepackage{graphicx}
\usepackage{textcomp}
\usepackage{subcaption}
\usepackage{xcolor}
\usepackage{soul}
\usepackage{url}
\usepackage{booktabs} 
\usepackage{authblk}
\providecommand{\keywords}[1]{\textbf{\textit{keywords---}} #1}
\newcommand{\beq}{\begin{equation}}
\newcommand{\bea}{\begin{eqnarray*}}
\newcommand{\eea}{\end{eqnarray*}}
\newcommand{\eeq}{\end{equation}}
\newcommand{\bdis}{\begin{displaymath}}
\newcommand{\edis}{\end{displaymath}}
\DeclareGraphicsExtensions{.pdf,.jpg,.jpeg,.png,.eps}
\def\BibTeX{{\rm B\kern-.05em{\sc i\kern-.025em b}\kern-.08em
    T\kern-.1667em\lower.7ex\hbox{E}\kern-.125emX}}
\begin{document}

\title{Optimising Parameters in Recurrence Quantification Analysis of Smart Energy Systems}

\author[1]{Georgios Giasemidis}
\author[2]{Danica Vukadinovi\'c Greetham}
\affil[1]{georgios@countinglab.co.uk, CountingLab Ltd., UK}
\affil[2]{danica.vukadinovic-greetham@open.ac.uk, Knowledge Media Institute, Open University, UK}
\date{}

\maketitle

\begin{abstract}
Recurrence Quantification Analysis (RQA) can help to detect significant events and phase transitions of a dynamical system, but choosing a suitable set of parameters is crucial for the success. From recurrence plots different RQA variables can be obtained and analysed. Currently, most of the methods for RQA radius optimisation are focusing on a single RQA variable. In this work we are proposing two new methods for radius optimisation that look for an optimum in the higher dimensional space of the RQA variables, therefore synchronously optimising across several variables. We illustrate our approach using two case studies: a well known Lorenz dynamical system, and  a time-series obtained from monitoring energy consumption of a small enterprise. Our case studies show that both methods result in plausible values and can be used to analyse energy data. 
\end{abstract}


\section{Introduction}
 Observing a particular dynamical system, defined by  human behaviour, often results in a complex, non-linear time-series. Recurrence, one of its fundamental properties that captures the system's phase space, is a  helpful tool to analyse the system. A recurrence plot \cite{eck87}, \cite{zbi92} can unfold the latent repeating patterns and is a visual tool  that facilitates  an investigation of the system.  Recurrence quantification analysis (RQA) helps to quantify the recurrent structures revealed by these plots, by computing  time dependent RQA variables  \cite{fab05}, \cite{mar13}. RQA variables can detect significant events and phase transitions, such as  financial crashes, transitions in climate systems, etc.

There are many RQA variables or measures in literature that have been used to detect important insights about dynamical systems.  The applications include heart rate\cite{zbi06}, financial time-series \cite{fab05}, etc. 
Only recently, RQA was applied to energy data  obtained from high resolution monitoring of electricity usage \cite{hat18} and was used to map different appliances, to detect faulty devices and identify unexpected usage patterns. 
The benefits of RQA include relatively simple calculations, powerful investigative characteristics, and applicability to a wide range of systems. One of the main challenges is to choose a set of parameters that are suitable for the application. Therefore, parameters need to be optimised according to the observed system's characteristics and the application. In \cite{mar11} several potential issues are recognised when choosing parameters. This is especially true with emerging areas of applications such as high resolution electric energy data.

The novelty of this work is twofold: (i) to the best of our knowledge, this is the first application of parameter optimisation for RQA to electricity data from Small and Medium Enterprises (SMEs), (ii) we introduce two practical methods for optimising the radius for RQA, which address the following open question in the literature. In existing work,  the radius is optimised for each RQA variable, and different RQA variables might result in different optimal radius for the same data. In addition, a decision is made after visual inspection of the plots of the different RQA variables. To remedy this, our methods use all five main RQA variables to automatically identify the optimal radius without further visual assessment.

In this work, we start with a quick overview of energy disaggregation literature in Section~\ref{sec:lr}. Fundamentals of RQA are given in Section \ref{sec:rp} and we list a selection of variables relevant for our energy case study in Section~\ref{sec:rqa-vars}. Our main contribution, two novel methods for radius optimisation are given in Section~ \ref{sec:met}. The methods are applied to two case studies, the Lorenz system and an energy data set in Section~\ref{sec:app} and we list the benefits and limitations of newly proposed methods in Section~\ref{sec:dis}.

\section{Previous work}
\label{sec:lr}
An area of extensive academic interest over the last decade, non-intrusive load monitoring (NILM) aims to identify electric load components (appliances) based on measured aggregated load. 

An overview of main concepts, techniques and algorithms of NILM  can be found in \cite{zei11, kle16}. The classical Hart's algorithm \cite{har92} starts by detecting significant changes in the signal and then clusters them in order to find  different  states' changes of each appliance.  

In the last $25$ years many different techniques were developed  using unsupervised  and supervised learning,  and low and high frequency sampling. While high frequency sampling could detect  fine features of the signal, most of smart meters work in low frequency mode - reporting usage at 1s or less frequently. Supervised learning algorithms contain a training phase, to create  a map of  all possible devices and their features.  After training phase, the disaggregation is performed using 
optimisation or pattern recognition. The disaggregation can be formulated as an optimisation problem, more precisely  error minimisation. In this problem, the known devices need to be  identified, such that their sum is closest to the actual consumption, usually requiring combinatorial optimisation techniques to find the solution. Alternatively, pattern recognition methods, classifiers, neural networks, hidden Markov Models and similar  are used to identify devices \cite{aia16}. 
In unsupervised learning, the training phase may happen in parallel with the disaggregation. Most of academic work in this area is concentrated on recognising household's appliances, and combinatorial optimisation and factorial Hidden Markov Models are currently regarded by most scientists as the state of art. 

Recent unsupervised approaches for households include using probabilistic graphical models to represent appliances, where the models' parameters are learned during training phase \cite{par12}. Both aggregate power reading and differences between two consecutive readings are used to create a version of difference Hidden Markov Model (HMM)  for each appliance.
Factorial HMMs, where several HMMs  are evolved independently and  some function of all the hidden states is the output, featured in \cite{zic12} obtaining good disaggregation results on home appliances. Another application of factorial HMM is given in \cite{aia16} where simultaneously running two devices was explicitly modelled as an interaction chain in HMM. Better disaggregation accuracy was obtained for devices with known interactions property while not much difference was observed for devices with unknown mutual interactions. 

In \cite{kel15} three deep neural network architectures were trained and tested on home appliances disaggregation. A recurrent neural network, so called long short term memory (LSTM) performed well on two-state appliances (on and off, e.g. toaster)  but not on multi-state  appliances that can be in many different discrete states (e.g. washing machine). While the preliminary results are promising, literally millions of trainable parameters represent a challenge.  Google recently reported that the deployment of an ensemble of deep neural networks trained on diverse data obtained from thousands of sensors reduced cooling data centre energy costs by $40\%$\footnote{\url{https://deepmind.com/blog/deepmind-ai-reduces-google-data-centre-cooling-bill-40/}}.

Other recently developed complex techniques that are comparable with factorial HMMs and combinatorial optimisation include using spatiotemporal pattern networks between several variables such as indoor and outdoor temperature and time of the day, in addition to whole building electricity and its moving average to obtain disaggregation \cite{liu18}.  

In \cite{ega15} factorial HMMs are combined with particle filtering,  to estimate disaggregated appliance states. Particle filtering is well suited to non-linear behaviour and non-Gaussian noise. 

While most of methods are developed and tested for households or the largest enterprises where economy of scale helps to justify investments, small and medium companies are somewhat left out
\cite{sch16}.

Recently, recurrence quantification analysis (RQA)  was applied to supervised disaggregation of small business load in \cite{hat18}. RQA,  suitable for non-linearity and complex time-dependencies, was combined with principal component analysis to create an easy visualisation of disaggregation for users.  Here we explore parameter optimisation of such approach with the aim to investigate a possibility of unsupervised system based on RQA in the future.

\section{Parameter optimisation}
\subsection{Recurrence Plot}
\label{sec:rp}
Recurrence plots capture the recurrent states of a complex system \cite{eck87}. A state of the system is considered recurrent when it is in a close neighbourhood of a previous state of the system in the phase space. Given a time-series of $n$ observations $X = \{x_1, \ldots, x_n\}$, the phase space is defined by transforming the readings into time-delayed vectors at each time-step, i.e.
$\textbf{Y}=\{Y_1, \ldots, Y_m\}$, where $Y_i = (x_i, x_{i-\tau}, \ldots, x_{i-(D-1\tau)}) \in \mathbb{R}^D$, $\tau$ is the delay and $D$ is the dimension of the phase-space, and $m=n-(D-1)\tau$.

The distance matrix $DM(i, j) = ||Y_i - Y_j ||_2$,  $i, j = 1, \ldots, m$, is the Euclidean distance between vectors $Y_i$ and $Y_j$ in the phase-space. The recurrence plot follows from the distance matrix, defined as
\beq
\label{eq:rp}
R(i,j)=H(\epsilon-DM(i,j)),
\eeq
where $H(x)$ is the Heaviside step function. 

The entry $R(i,j)$ equals one and the states $Y_i$ and $Y_j$ are considered recurrent, when the distance between $Y_i$ and $Y_j$ is within an $\epsilon$-radius in phase-space. As  a state $Y_i$ in phase space corresponds to a time-step of the original time-series $X$, recurrence plots inform us of recurring patterns within our current time-series.

From the aforementioned definitions, it becomes evident that the recurrence plot depends on three parameters, the delay $\tau$, the embedding dimension $D$ and the radius $\epsilon$, to capture the correct dynamics of a system with noise. 
 
\subsection{RQA variables}
\label{sec:rqa-vars}
Following \cite{hat18}, we concentrate on several RQA variables, all obtained from the recurrence plot matrix $R$: 
\begin{itemize}
\item {\bf REC}, the percentage of all points within the square window of size $W$ that are recurrent.
\item {\bf DET}, the percentage of the recurrence points that form line segments parallel to the matrix diagonal.  
 \item {\bf  ENT}, the Shannon entropy of the distribution of diagonal line segments.
 \item {\bf  LAM}, the percentage of the recurrence points that form vertical line segments.  
\item {\bf TT}, the average length of the vertical line segments,
\end{itemize}
These variables are frequently used across different applications \cite{mar13, fab05, vla12}. DET  relates to repeating or deterministic patterns within the system, ENT represents complexity,  where small and large ENT signify periodic and unpredictable behaviour respectively and LAM represents stationary behaviour.

\subsection{Methodology}
\label{sec:met}
Several studies have focused on the optimisation of the three parameters \cite{DING20081457, mar11}. Common methods for optimising the delay include the first minimum of either the autocorrelation function or the mutual information of the time-series \cite{kan03, mar11}. Here, we choose the latter as it captures non-linear characteristics of the time-series. The embedding dimension is often determined using the false nearest neighbours parameters \cite{Kennel1992}. In this study, we use a modification of this algorithm, introduced by Cao \cite{CAO199743}, in order to be parameter-free and efficient.

Several methods have been proposed for choosing the radius. A too small value allows no recurrent patterns, a too large values may result in false recurrences \cite[references therein]{DING20081457}. Several rules of thumb have been proposed, such as a value corresponding to 1\% of REC \cite{zbi02}, $\epsilon=0.1\sigma$, where $\sigma$ is the standard deviation of the time-series, a value that does not exceed the 10\% of the mean or the maximum of the phase-space \cite[references therein]{DING20081457, Schinkel2008}. 

Other more sophisticated methods have been proposed in \cite{DING20081457,YANG2015361, gao09}. Although these methods are distinct, they share some common features. They create a surrogate signal, which is the original signal with additive noise. Then they run RQA for several segments of the original and the surrogate signals. The objective is to identify the radius that is the best in discriminating between the original and the surrogates dynamics. This is achieved by optimising several measures such as the average loss of decision action \cite{DING20081457}, the area under the ROC \cite{Schinkel2008} or the quality loss function \cite{YANG2015361}. However, these scores are applied per RQA variable, and in most cases, the score is optimised at different radii values for different RQA variables. Therefore, the question remains, how can one decide in a practical way which radius to choose, as further analysis is required to decide which RQA variable is more significant \cite{Schinkel2008}. Here, we propose two practical methods to identify the optimal radius without having to choose `the most important' variable.

\subsubsection{Method 1}
In the first method, a surrogate signal is produced. A number $N$ of segments are randomly drawn from the original time-series and its surrogate. For several values of the radius, RQA variables are computed for each segment of the original and surrogate signals. These correspond to points $Q = (REC, DET, ENT, LAM, TT) \in \mathbb{R}^5$ in the 5-dimensional RQA space (the space of the RQA variables defined in the previous section). If we label the points corresponding to the original signal as ``Cluster 1'' and those corresponding to the surrogate as ``Cluster 2'', as the radius is varied, these points in each cluster spread or cluster together depending the quality of the clustering. To assess the quality of the clustering we use scores from the Machine Learning clustering literature. The inertia, the within-cluster distance, is defined as
\beq
I = \sum_i \min _{\mu_j \in C} || Q_i - \mu_j||_2,
\eeq
where $\mu_j \in \mathbb{R}^5$ is the centroid of the cluster $j$. The inter-cluster distance is defined
\beq
d_C = || \mu_1 - \mu _2 ||_2
\eeq 
The maximum discrimination between the original and surrogate signals is achieved when the inertia is minimised and inter-cluster distance of the centres of the clusters is maximised. In order for the two measures to be of comparable sizes when combined, they are normalised to take values in the interval $[0, 1]$. The optimal radius is given by
\beq
\epsilon ^{*} = \arg\min_{\epsilon} (\hat{I}(\epsilon) - \hat{d}_C(\epsilon)),
\label{eq:method1}
\eeq
where $\hat{I}$ and $\hat{d}_C$ are the normalised inertia and inter-cluster distance respectively. 

\subsubsection{Method 2}
In the second method, the RQA variables of the original signal are computed for different values of the radius, $Q_{\text{or}}(\epsilon) \in \mathbb{R}^5$. These points form a curve in the 5-dimensional phase-space. A sample of surrogates is also produced for each value of the radius, their RQA variables are extracted and their mean value is computed, $Q_{\text{su}}(\epsilon) \in \mathbb{R}^5$. The maximum discrimination occurs when the Euclidean distance between the RQA variables of the original and mean surrogate signals is maximal, i.e.
\beq
\epsilon ^{*} = \arg\max_{\epsilon} ||Q_{\text{or}}(\epsilon) - Q_{\text{su}}(\epsilon)||_2.
\label{eq:method2}
\eeq 

Several surrogate functions have been used, such as shuffling or Fourier based surrogates \cite{DING20081457}. Here, we use an additive Gaussian noise to produce the surrogate \cite{Schinkel2008}, i.e. $X_{\text{wn}} = X + \mathcal{N}(0, \alpha \sigma)$, where $\sigma$ is the standard deviation of the original time-series $X$ and $\alpha$ controls the amount of noise.

\section{Applications}
\label{sec:app}
We apply the proposed methods to two systems, the Lorenz attractor and to real data from electricity demand from an SME \cite{hat18}. For all systems, the time-delay is identified through the first local minimum of the mutual information and the embedding dimension using Cao's algorithm \cite{CAO199743}. Then, we identify the radii values that correspond to a range of REC values of the system from 0\% to 100\% in increments of 0.5\%.
For the white-noise surrogate we use $\alpha=0.2$. 
In the following examples, for comparison, we also plot the values of the radius that correspond to $1\%$ REC, $0.1 \sigma$, $10\%$ of maximum and mean distance in the phase space.

\subsection{Lorenz Attractor}
The Lorentz system is a dynamic system of three differential equations \cite{lor63} resulting in chaotic behaviour for some choices of parameters and initial conditions:

\beq
\left(\frac{dx}{dt}, \frac{dy}{dt}, \frac{dz}{dt}\right) =(\sigma{y-x}, x(\rho -z)-y, xy-\beta z)
\eeq

We create $N=3000$ times-steps of the system with step-size $ 0.01$ and parameters $\sigma=10, \rho=28, \beta=8/3$ that result in chaotic behaviour. The time-delay is determined to be $\tau=17$ and the embedding dimension $D=3$, consistent to the literature \cite{CAO199743,Kennel1992}.

\begin{figure}
	\centering
	\begin{subfigure}[t]{0.45\textwidth}
		\centering
		\includegraphics[width=\linewidth]{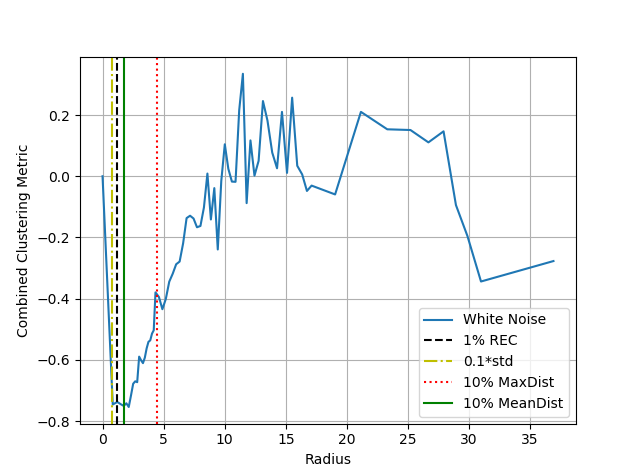}
		\caption{The metric based on clustering of the RQA values of the segments of the original signal and the white noise (blue).}
		\label{fig:lorenz-clustering}
	\end{subfigure}
	~ 
	\begin{subfigure}[t]{0.45\textwidth}
		\centering
		\includegraphics[width=\linewidth]{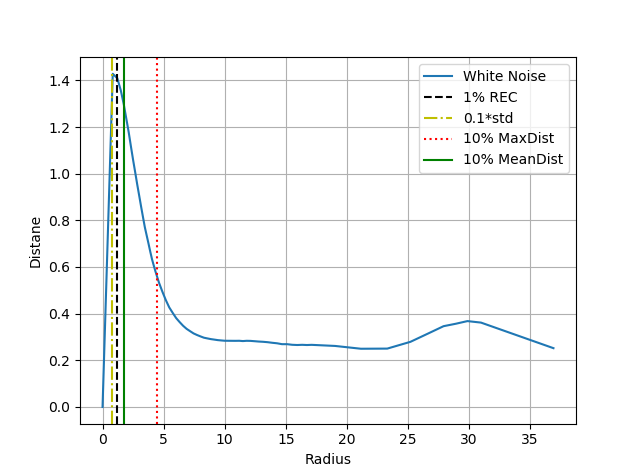}
		\caption{The distance measure in the 5-dimensional space between the original signal and the white noise (blue).}
		\label{fig:lorenz-distances}
	\end{subfigure}
	\caption{Lorenz attractor}
	\label{fig:lorenz}
\end{figure}

Figure \ref{fig:lorenz-clustering} shows the measure \eqref{eq:method1} which assesses the quality of the clustering of the RQA values between the original and the surrogate signals. The maximum discrimination (i.e. minimum value) to the white-noise surrogate is achieved at the value $\epsilon^* = 2.14$, which is comparable to the optimal value in \cite{Graben2013,Graben2016}, and corresponds to $3\%$ REC, a conclusion similar to \cite{DING20081457,YANG2015361} too.

The distance in the RQA-space between the original signal and its surrogate as a function of the radius is presented in Figure \ref{fig:lorenz-distances}. 
The optimal value is at at $\epsilon^* = 0.84$, corresponding to $0.5\%$ REC, which also coincides with $0.1 \sigma$ rule of thumb.

\subsection{Case study: energy data}

\begin{figure}
	\centering
	\includegraphics[width=\linewidth]{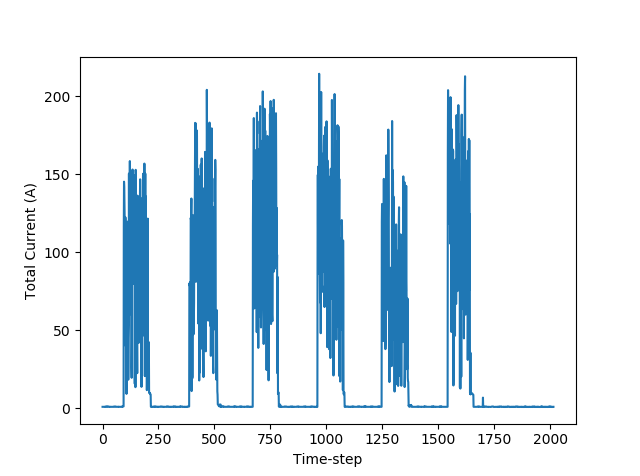}
	\caption{A weekly profile of the total current from a dry cleaners business. The business is closed on Sundays.}
	\label{fig:weekly-profile}
\end{figure}

The energy data corresponds to total electric current at five minute resolution from a dry cleaners, a small UK business \cite{hat18}, for a period of 6 weeks from 11-th of September to 22-nd of October 2017. Figure \ref{fig:weekly-profile} shows a typical weekly profile. 

As we are interested in the dynamics while the business is open, we include only the operational times in our analysis. To identify the operational times, we searched for the maximum reading in the data between 10pm and 6am at nigh, where the business is closed, see Figure \ref{fig:weekly-profile}. The maximum value is $1.4A$ and any values greater than this are treated as ``on'' states, i.e. operational times. 

\begin{figure}
	\centering
	\begin{subfigure}[t]{0.45\textwidth}
		\centering
		\includegraphics[width=\linewidth]{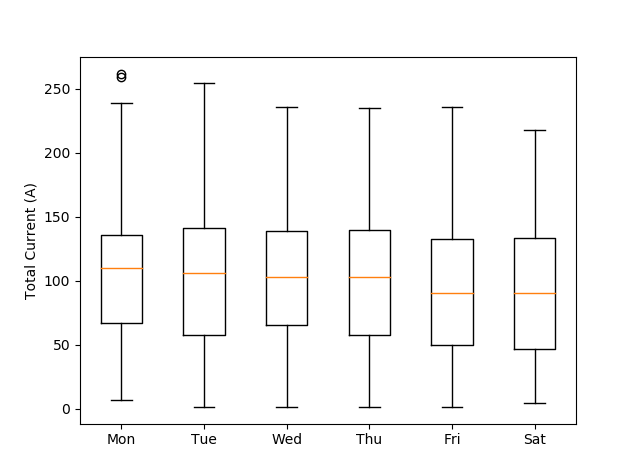}
		\caption{Day of the week (the business is closed on Sundays)}
		\label{fig:boxplot-dow}
	\end{subfigure}
	~ 
	\begin{subfigure}[t]{0.45\textwidth}
		\centering
		\includegraphics[width=\linewidth]{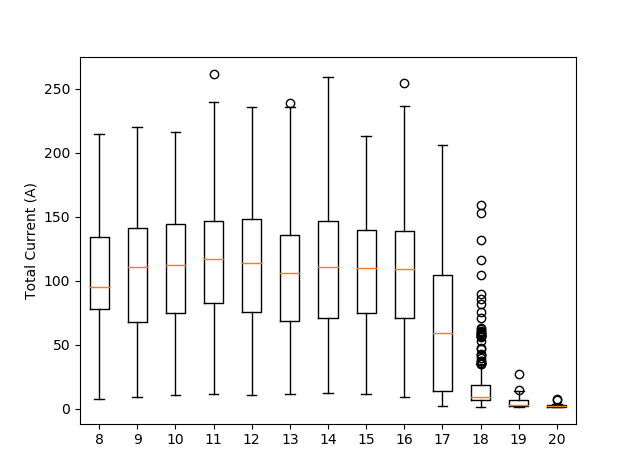}
		\caption{Hour of the day (business is closed before 8am and after 9pm).}
		\label{fig:boxplot-tod}
	\end{subfigure}
	\caption{Distribution of electrical current readings.}
	\label{fig:boxplots}
\end{figure}

In Figure \ref{fig:boxplot-dow}, we plot the distribution of readings for each day of the week. We observe that all days have similar medians and quartiles. 
In Figure \ref{fig:boxplot-tod}, the distribution of readings for each hour of the day is plotted. The distributions of the readings are similar for the hours between 8am to 5pm, the current usage drops between 5pm and 6pm, however after 6pm the median and quaritles of the electrical current are much lower, but higher than the ``off-state''. This is because most devices have been turned off and not in use, whereas one or two devices still operate until the closure of the business.

%

From Figure \ref{fig:boxplots}, we observe no daily or hourly patterns. This is expected as this is data from an SME, which operates in full demand during its operational hours, in contrast to household data which has daily and weekly behavioural patterns. The lack of strong periodicity patterns and the high standard deviation of the distributions indicate that the readings have a strong stochastic component, which is also unveiled from our analysis below.

\begin{figure}
	\centering
	\begin{subfigure}[t]{0.45\textwidth}
		\centering
		\includegraphics[width=\linewidth]{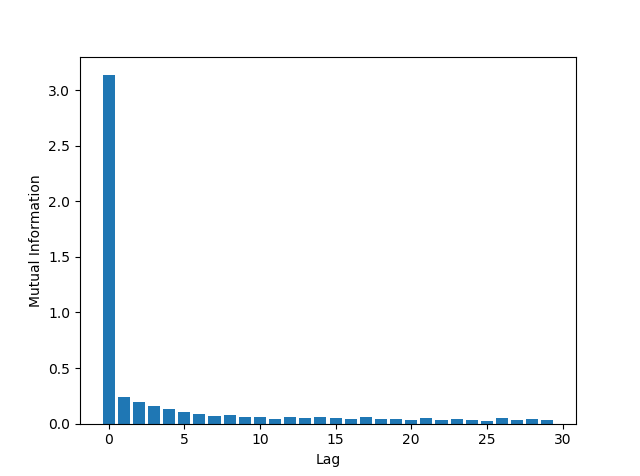}
		\caption{Mutual information for dry-cleaners data.}
		\label{fig:mutual-information}
	\end{subfigure}
	~ 
	\begin{subfigure}[t]{0.45\textwidth}
		\centering
		\includegraphics[width=\linewidth]{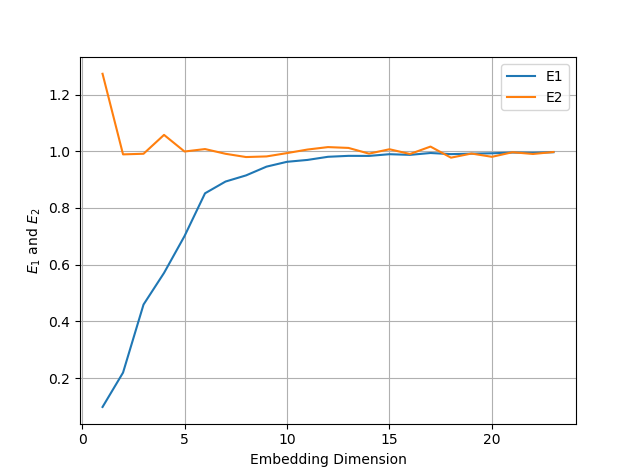}
		\caption{Cao's metrics for dry-cleaners.}
		\label{fig:embedding-dimension}
	\end{subfigure}
	\caption{Mutual information and embedding dimension for the dry-cleaners.}
	\label{fig:mi-d}
\end{figure}

For the optimisation of the RQA parameters we focus on a period of two weeks from 11-th to 25-th of September 2017. In Figure \ref{fig:mutual-information} we plot the mutual information for several lags. There is a steep drop in the mutual information for lags greater than 1, with a constant plateau, which indicates that the optimal value is $\tau=1$. Figure \ref{fig:embedding-dimension} demonstrates how Cao's measures $E_1$ and $E_2$ \cite{CAO199743} change with the embedding dimension. The fluctuation of $E_2$ around the value 1 and the slow increase of measure $E_1$ to high values of embedding dimension are characteristic of signals with a dominant stochastic component \cite{CAO199743,Kennel1992}, which is indeed the case for real load data. We conclude that the embedding dimension should be $D=10$.

\begin{figure*}
	\centering
	\begin{subfigure}[t]{0.48\textwidth}
		\centering
		\includegraphics[width=\linewidth]{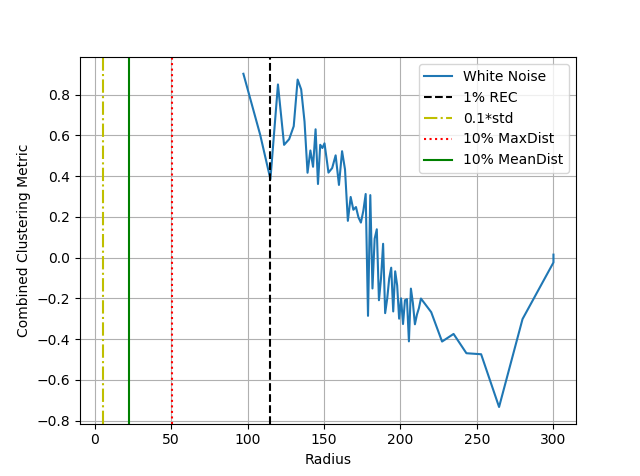}
		\caption{The metric based on clustering of the RQA values of the segments of the original signal and the white-noise surrogate (blue).}
		\label{fig:signal-clustering-dim10}
	\end{subfigure}
	~ 
	\begin{subfigure}[t]{0.48\textwidth}
		\centering
		\includegraphics[width=\linewidth]{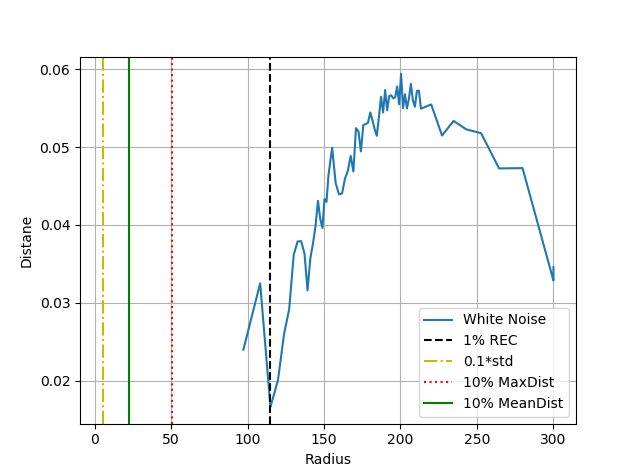}
		\caption{The distance measure in the 5-dimensional space between the original signal and the white-noise surrogate (blue).}
		\label{fig:signal-distances-dim10}
	\end{subfigure}
	\caption{Dry cleaners ($\tau=1$ and $D=10$).}
	\label{fig:signal-dim10}
\end{figure*}

Having determined the time delay and embedding dimension, we use these values to estimate the radius of RQA. Figures \ref{fig:signal-clustering-dim10} and \ref{fig:signal-distances-dim10} show how the radius changes for methods 1 and 2, respectively. We observe that the radius extends to high values, which is expected as the embedding dimension of the system is relatively large. The optimal values are at $\epsilon ^* = 264.5$ ($75\%$ REC) and $\epsilon ^* = 200.5$ ($35\%$ REC). These optimal values occur at high percentages of the REC, results that are consistent with the findings in \cite{DING20081457}. In this study, the authors studied heart rate dynamics data (which consists of linear, chaotic and stochastic components) at different embedding dimensions and they found that as the embedding dimension increases, the optimal radius (according to their method) occurs at higher levels of REC. Particularly, at $D=10$, their optimal radius occurs at $68\%$ REC, which is consistent with our findings. 

\begin{figure*}
	\centering
	\begin{subfigure}[t]{0.48\textwidth}
		\centering
		\includegraphics[width=\linewidth]{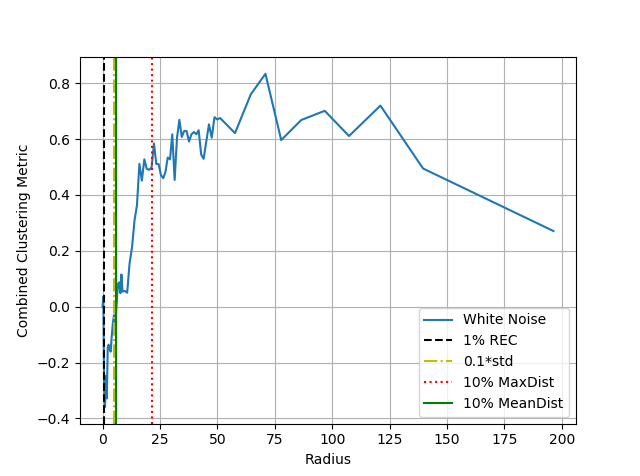}
		\caption{The metric based on clustering of the RQA values of the segments of the original signal and the white-noise surrogate (blue).}
		\label{fig:signal-clustering}
	\end{subfigure}
	~ 
	\begin{subfigure}[t]{0.48\textwidth}
		\centering
		\includegraphics[width=\linewidth]{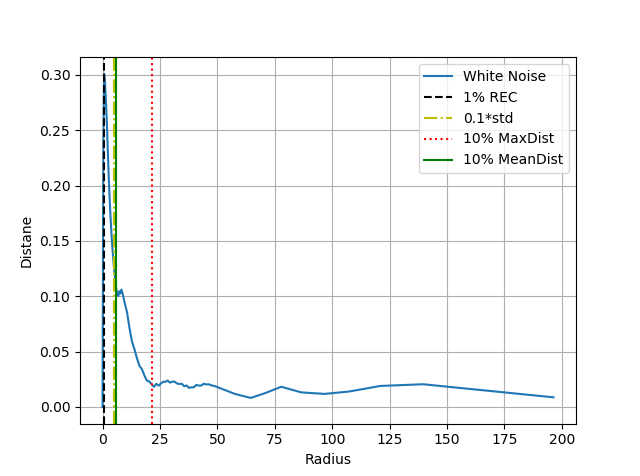}
		\caption{The distance measure in the 5-dimensional space between the original signal and the white-noise surrogate (blue).}
		\label{fig:signal-distances}
	\end{subfigure}
	\caption{Dry cleaners ($\tau=1$ and $D=1$).}
	\label{fig:signal}
\end{figure*}

In order to further understand the dependence of the optimal radius with the embedding dimension, we apply the proposed methods for $\tau=1$ and $D=1$. Figure \ref{fig:signal-clustering} shows how the clustering measure eq. \eqref{eq:method1} varies with the radius. We observe that the radius extends to very high values in order to span REC values from $0\%$ to $100\%$. The white-noise surrogate achieves a maximum discrimination at $\epsilon^* = 1.08$ which corresponds to $1.5\%$ REC. Figure \ref{fig:signal-distances} shows that the distance measure eq. \eqref{eq:method2} is maximised at $\epsilon^* = 0.69$ ($1\%$ REC). 
The discrimination between the original signal and the surrogate reduces with the increase of the radius as expected, due to the identification of false recurrences in the surrogate. For $D=1$, we observe that optimal values occur at lower values of the REC levels as in \cite{DING20081457}.

\section{Discussion}
\label{sec:dis}
Optimising the parameters in RQA is a non-trivial task and the success of the proposed methods depends on the applied system \cite{mar11}. In this study, we apply existing and novel methods of RQA parameter optimisation to smart energy systems. The time delay and embedding dimension are successfully determined through the mutual information and Cao's algorithm respectively. For the optimal radius, we propose two new methods, based on the simultaneous discrimination power of several RQA variables between the original signal and a noise surrogate. 

We validate our results using a well-studied Lorenz system, where our findings are consistent with the previous literature. Our main interest is in the new application, energy usage data, where both methods 1 and 2 behave qualitatively as expected, producing similar results.  The discrimination power reaches its maximum value at a relative low value of the radius and the REC percentage and then decreases as the radius increases due to the occurrence of false recurrences in the surrogates. Note that the optimal radius value depends on the embedding dimension, as expected. We leave further research to better understand this dependence for future studies.

\section*{Acknowledgment}
This work was carried out with the support of BEIS. We thank our project partner AND Technology Research \\(\url{http://andtr.com/}) for providing us with the data.

\bibliographystyle{IEEEtran}

\bibliography{bib}

\end{document}